%% file: manuscript.tex
\documentclass[12pt]{article}
\usepackage{amsmath}
\usepackage{graphicx,psfrag,epsf}
\usepackage{enumerate}
\usepackage[round]{natbib}
\usepackage{url} 

\include{preamble}

\newcommand{\blind}{0}

\begin{document}

\def\spacingset#1{\renewcommand{\baselinestretch}%
{#1}\small\normalsize} \spacingset{1}

\makeatletter
\newcommand{\setword}[2]{%
  \phantomsection
  #1\def\@currentlabel{\unexpanded{#1}}\label{#2}%
}
\makeatother


\if0\blind
{
  \title{\bf Harmonizing Fully Optimal Designs with Classic Randomization in Fixed Trial Experiments}
\author{
Adam Kapelner,\\
Department of Mathematics, Queens College, CUNY,\\
~\\
Abba M. Krieger,\\
Department of Statistics, The Wharton School of the University of Pennsylvania,\\
~\\
Uri Shalit and David Azriel\\
Faculty of Industrial Engineering and Management, The Technion}
  \maketitle
} \fi

\if1\blind
{
  \bigskip
  \bigskip
  \bigskip
  \begin{center}
    {\LARGE\bf Harmonizing Fully Optimal Designs with Classic Randomization in Fixed Trial Experiments}
\end{center}
  \medskip
} \fi

\bigskip
\begin{abstract}
There is a movement in design of experiments away from the classic randomization put forward by Fisher, Cochran and others to one based on optimization. In fixed-sample trials comparing two groups, measurements of subjects are known in advance and subjects can be divided optimally into two groups based on a criterion of homogeneity or \qu{imbalance} between the two groups. These designs are far from random. This paper seeks to understand the benefits and the costs over classic randomization in the context of different performance criterions such as Efron's worst-case analysis. In the criterion that we motivate, randomization beats optimization. However, the optimal design is shown to lie between these two extremes. Much-needed further work will provide a procedure to find this optimal designs in different scenarios in practice. Until then, it is best to randomize.
\end{abstract}

\noindent%
\vfill

\newpage

\section{Introduction}

In this short survey, we wish to investigate performance differences between completely random experimental designs and non-random designs that optimize for observed covariate imbalance. We demonstrate that depending on how we wish to evaluate our estimator, the optimal strategy will change. We motivate a performance criterion that when applied, does not crown either as the better choice, but a design that is a harmony between the two of them. We demonstrate our claim through simulation and a heuristic argument. Our observations open the door to future fundamental research on this age-old debate.

\subsection{Background}

We consider a classic problem: a two-arm, fixed, non-sequential experiment whose goal is to estimate and test the treatment effect. This experiment has a clearly defined \emph{outcome} of interest (also called the \emph{response} or \emph{endpoint}) and we scope our discussion to the response being continuous and uncensored.

Synonymously referred to as a \emph{design}, a \emph{randomization}, an \emph{allocation} or an \emph{assignment} and constructed via a \emph{strategy}, \emph{algorithm}, \emph{method} or \emph{procedure} is the division of $n$ \emph{individuals} (\emph{subjects}, \emph{participants} or \emph{units}) into a treatment group and a control group ($T$ and $C$), the two \emph{arms}. Historically, standard Bernoulli draws for each individual is termed \emph{complete randomization} and is sometimes called the \qu{gold standard}. Any other design is termed a \emph{restricted randomization} because it is a restricted to a subset of all possible allocations. 

Why has complete randomization been given this high distinction? \citet[p. 245]{Cornfield1959} gives two reasons: (1) known and unknown covariate differences among the groups will be small and bounded and (2) it forms a basis for inference. However, there is a large problem which was identified at the inception of experimentation: sometimes differences are exhibited in the distribution of observed covariates among individuals within the two groups under some of the assignments produced by complete randomization. The amount of covariate difference we term \emph{observed imbalance}. Through an abuse of terminology, the literature denotes this as simply \emph{imbalance}, but this is an ambiguous term since it usually ignores the state of imbalance in the unobserved covariates. Observed imbalance is summed up in a numerical metric that can be defined a number of different ways. By convention we consider \qu{larger} imbalance values as worse.

Mitigating the chance of large observed imbalances is the predominant reason for a priori restrictions on the randomized allocation. \citet[p. 251]{Fisher1925} wrote \qu{it is still possible to eliminate much of the \ldots heterogeneity, and so increase the accuracy of our [estimator], by laying restrictions on the order in which the strips are arranged}. Here, he introduced \emph{blocking}, a restricted design still popular today. \citet[p. 366]{Student1938} wrote that after an unlucky, highly imbalanced randomization, \qu{it would be pedantic to [run the experiment] with [an allocation] known beforehand to be likely to lead to a misleading conclusion}. His solution is for \qu{common sense [to prevail] and chance [be] invoked a second time}. In doing this \emph{rerandomization}, all allocations above a predetermined threshold of observed imbalance are eliminated, a strategy that has been rigorously investigated only recently \citep{Morgan2012, Li2016}. Another idea is to allocate treatment and control among similar subjects by using a matching algorithm \citep{Greevy2004}, a tool by and large popularized in the field of observational studies to minimize confounding. Additionally, once the imbalance metric is explicitly defined, one can formulate the procedure as a binary integer programming problem, a form of numerical optimization \citep{Bertsimas2015, Kallus2018}. This is frequently solved via branch and bound \citep{Land1960} which can provide near optimal designs. One can also employ other heuristics that optimize but simultaneously preserve randomness \citep[e.g.][]{Krieger2016}.

The above gives a short introduction to the wide range of design ideas. At the \qu{extremes} of this range, there is complete randomization and the optimized observed imbalance design. We now seek to compare them. We begin with a simple, intuition-building scenario where complete randomization provides superior performance over the optimized imbalance design.

\subsection{A Simple Illustration and the Efron Bound}\label{subsec:example_tradeoff}

Consider a fixed trial experiment with $n$ subjects comparing a treatment group to a control group. There is one known covariate vector $\x$ and one unknown covariate vector $\z$ and a population average treatment effect (PATE) of 1 which we denote $\beta_T$. Consider the additive treatment effect model, $\y = \beta_T  \w + \x + \z$ where $\w$ denotes the allocation vector with entries +1 for treatment or -1 for control. 

For the purposes of illustration, consider the scenario where the subjects are grouped into $m = n / 2$ pairs where the two subjects within each pair share the same $x$ value but differ by a $z$ value denoted $a$ and $-a$. The value of the observed covariate is the same for the two observations in any pair. The value in pair $i$ is $\delta(i-\frac{m+1}{2})$. We will see shortly that this structure is the most adversarial if we optimize for observed imbalance.


We now wish to compare completely randomized design to optimal design. To follow assumptions introduced in the next sections, we will limit our discussion here to \emph{forced balance designs} \citep[chap 3.3]{Rosenberger2016} where the treatment group and the control group are both coerced to have equal numbers of subjects.

Considering the standard imbalance metric of Mahalanobis distance between the sample average measurements in the treatment group and the sample average measurements in the control group,  the optimal in this case can be found by a priori matching \citep{Greevy2004}. In this case, the optimal design is to match the two people in each pair. The member of the pair that is given the treatment is randomly chosen. 

In this matched pairs allocation procedure, there are only $2^3 = 8$ allocations, a restricted subset of allocations under complete randomization with forced balance (CRFB) where there are $\binom{6}{3} = 20$ possible choices. Parenthetically, the optimal design in this case will have many allocation vectors but if the covariates were truly continuous, there would be only one unique optimal partition. 




We then employ the simple differences-in-means estimator, $\betaThat := (\bar{Y}_T - \bar{Y}_C) / 2$ where the division by two is only due to the fact that $\w$ has entries -1 and +1 (not 0 and 1). The criterion for the performance of $\betaThat$ under the different designs we choose to be mean squared error (MSE), an average over all possible experimental replicates given these subjects with their particular $\x$ and $\z$ constant.

The ratio of the standard deviation of the observed covariate to the standard deviation of the unobserved covariate $\eta$ is a pivotal quantity in this illustration. Using this equivalent parameterization, $\delta=\frac{a\sqrt{12}}{\sqrt{m^2-1}} \eta$. Tab.~\ref{tab:simple_example_results_general} shows (a) the mean observed imbalance as measured by the squared difference in averages of the two groups i.e. $(\bar{x}_T - \bar{x}_C)^2$, (b) the mean unobserved imbalance i.e. $(\bar{z}_T - \bar{z}_C)^2$ and (c) the MSE of the difference-in-means estimator for both the restricted and CRFB procedures. 

\begin{table}[ht]
\centering
\caption{Metrics for a general adversarial example for two designs.}
\begin{tabular}{cccc}
  \hline
 & mean observed & mean unobserved & MSE of the \\ 
 & imbalance  & imbalance & treatment estimator \\ 
  \hline
random (CRFB) & $\frac{4a^2 \eta^2}{2m-1}$ & $\frac{4a^2}{2m-1}$ & $\frac{a^2(\eta^2+1)}{2m-1}$ \\ 
restricted (matching) & 0 & $\frac{4a^2}{m}$& $\frac{a^2}{m}$ \\ 
   \hline
\end{tabular}
\label{tab:simple_example_results_general}
\end{table}

The problem is calibrated so that when $\eta = 1$, the observed and unobserved variables carry the same weight in determining the response. Note that the smaller the $\eta$ (i.e., the less important the observed covariate), the more our estimator favors randomization over matching. In fact, randomization is preferred so long as $\eta < \sqrt{(m-1)/m}.$ For example, the case of $m=3$, $a=1.5$ and $\delta = 1$ (implying $\eta \approx 0.54 < \sqrt{2/3} \approx 0.82$) is shown below in tab.~\ref{tab:simple_example_results}.

\begin{table}[ht]
\centering
\caption{Results for a specific adversarial example.}
\begin{tabular}{cccc}
\hline
& mean observed & mean unobserved & MSE of the \\ 
& imbalance  & imbalance & treatment estimator \\ 
\hline
random (CRFB) & 0.53 & 1.80 & 0.58 \\ 
restricted (matching) & 0.00 & 3.00 & 0.75 \\ 
\hline
\end{tabular}
\label{tab:simple_example_results}
\end{table}

We observe that (1) the mean observed imbalance in the optimally designed experiments is zero, as expected. Any imbalance metric can be chosen as the distribution of covariate in the treatment group and control group are identical (2) the optimal imbalance procedure has worse imbalance on the unobserved covariate $z$ than the CRFB procedure. Further, (3) the estimation accuracy is worse under the optimal imbalance procedure even though the observed imbalance is optimal. Why is the variance of the treatment effect lower under randomization than matching  when $f$  is small? Because that is when $z$ is more important in determining $y$ and it is intuitive that imbalancing $z$ will adversely affect performance. Note that (2) and (3) illustrate the first reason for randomization provided by \citet{Cornfield1959} we quoted above.

The question is how to formalize the harm due to imbalancing the unobserved covariates into a criterion. One of the first attempts was introduced by \citet[sec. 5]{Efron1971}. He explains that when testing the null of the average treatment effect being zero, there is an inferential penalty incurred when the \qu{accidental bias}, defined as $\w^\top \z$, is non-zero (these expressions are generalized and explained in detail the next section). When considering replications of the experiments with different $\w$'s (the set of which are defined by the procedure), he derives the increase in variance of the simple mean differences treatment effect estimator to be $\z^\top \bSigmaw \z$ where $\bSigmaw := \var{\w}$, the variance-covariance matrix of the distribution of the allocation vectors from the procedure. 

A trivial bound then follows: its maximum cannot exceed the largest eigenvalue of $\bSigmaw$ for $\z$ normalized, i.e. 

\bneqn\label{eq:efron_bound}
\z^\top \bSigmaw \z \leq \lambda_{max} \normsq{\z}
\eneqn

(ibid, Equation 5.4, \citealp[p. 320]{Lachin1988} and \citealp[chap 4]{Rosenberger2016}). Thus, the worst case $\z$ is when it is the eigenvector corresponding to the largest eigenvalue (modulo a scaling), and this is exactly the $\z$ we have adversarially demonstrated in this example.

\section{Some Different Criterions}\label{sec:criterions}

The example of the previous section is compelling and possibly a justification for randomization in and of itself. However, it is cynically adversarial as such a $\z$ is extremely unlikely to occur in practice. Imagine a fixed $\x$ but a random $\z$ (e.g. the standard assumption of \emph{iid} sampling of the $z_i$'s). A realization of $\z$ that is non-trivially parallel to the worst eigenvector of $\bSigmaw$ (an arbitrary direction in $\reals^n$) is a low probability event whose probability shrinks exponentially as $n$ increases.

This criterion thus may not be the right choice for a practitioner. We now explore different criterions. But first we make clear our problem setup and assumptions.

\subsection{Problem Setup}

Assume a trial with a fixed number of subjects $n$ where each subject has $p$ fixed observed measurements ascertained before the study begins (assumption \setword{A1}{assumption:fixed_X}). We denote $\X \in \reals^{n \times p}$ as the measurements where the $i$th row represents the covariate of subject $i$, and it is denoted by $\x_i$, $i=1,\ldots,n$. All results and expressions to follow are conditional on $\X$ and thus the notation is dropped. 

The experiment begins when one allocation, a vector of manipulations denoted by $\w \in \allocspace := \braces{-1,+1}^n$, is administered to all subjects. The allocation is drawn from the \emph{design space}, denoted $\allocspace_0 \subset \mathcal{W}$, where the restriction is most made on the basis of $\X$. We assume that allocations have the \qu{mirror property} where treatments and controls can be switched with equal probability: $\prob{\w=\w_0} = \prob{\w=-\w_0}$ for all $\w_0 \in \mathcal{W}$ (assumption \setword{A2}{assumption:mirror}). 

The experiment ends when we assess the outcome $\y \in \reals^n$. We assume an additive effect model, i.e. $\cexpe{y_i}{\x_i,w_i}=\beta_T w_i + f_i$, $i=1,\dots,n$, where $\beta_T$ is the treatment effect and $f_i=f(\x_i)$ is some unknown function (assumption \setword{A3}{assumption:additivity}). Defining, $\z = \y - \cexpe{y_i}{\x_i,w_i}$; we obtain our main model, 

\bneqn\label{eq:main_model}
\y = \beta_T \w + \bv{f} + \z.
\eneqn

By the law of iterated expectation, $\z$ is mean-centered. 

We employ the simple mean differences estimator $\betaThat=\oneover{n} \w^\top \y$ to infer $\beta_T$. Sec.~\ref{pf:unbiasedness} proves that $\betaThat$ is unbiased. 

\subsection{A Worst-Case Criterion}\label{subsec:worse_case_criterion}

Conditional on a given realization of $\z$, the variance, equal to the mean squared error, can be expressed as:

\bneqn\label{eq:conditional_mse}
\cmsesubnostr{\w}{\betaThat}{\z} = (\f + \z)^\top \bSigmaw (\f + \z)
\eneqn

according to sec.~\ref{deriv:cond_var}, where $\bSigmaw := \expe{\w \w^\top}$, the variance-covariance matrix for $\w \in \allocspace_0$. This criterion represents the average error for a design, where the design is specified by the sufficient parameter $\bSigmaw$, conditional on one set of subjects (one $\z$).

Consider the minimax optimal design, one that minimizes the MSE based on the worst $\z$ i.e.

\beqn
\bSigmaw^* := \inf_{\bSigmaw \in \Omega_\Sigma} \sup_{\substack{||\z||^2 \le 2||\f||^2, \\ \z \in \reals^n}} \cmsesubnostr{\w}{\betaThat}{\z}
\eeqn 

where the condition of $\z$ being bounded is required to avoid a trivial infinity (the specific bound used is needed for the result below to hold). The design space $\Omega_\Sigma$ is defined to be all variance-covariance matrices of a generalized $n$-dimensional Bernoulli distributions. The worst $\z$ is where $\f+ \z$ is the scaled eigenvector corresponding to the largest eigenvalue of $\bSigmaw$ denoted $\lambda_{max}$ i.e. when Efron's bound is tight (eq.~\ref{eq:efron_bound}). Specifically, if we choose $\z=\f-\alpha \v_{max}$, where $\v_{max}$ is the eigenvector corresponding to the largest eigenvalue and $\alpha=||\f||^2/(1+2||\f||)$ then $||\z||^2 = 2||\f||^2$ (i.e., $\z$ belongs to the set over which the supremum is taken)  and $\f+\z=\alpha \v_{max}$ (i.e., it is the scaled eigenvector corresponding to the largest eigenvalue). Note that there is nothing special about the constant 2, it can be replaced by any number greater than 1.
 
When taking the infimum over designs, we can only minimize $\lambda_{max}$ as $\f$ and $\z$ are fixed. The solution is complete randomization where $\bSigmaw = \I_n$ and $\lambda_{max} = 1$ \citep[see][Problem 4.3]{Rosenberger2016}. Thus complete randomization would be minimax optimal for a specific $\x$ and $\z$ even if $\f$ is unknown.

This result is similar to the \qu{no free lunch} theorem proved by \citet[sec. 2.1]{Kallus2018}. Here, there is no free lunch in the sense that any design that balances observed covariates via a restriction of the sample space $\allocspace_0$ can inadvertently trigger higher variance in $\betaThat$ due to adversarial unobserved covariates $\z$. Knowledge of $f$ does not provide assistance.

\subsection{A Mean Criterion}\label{subsec:mean_criterion}

Conditioning on a single $\z$ is a limiting assumption since there are infinitely many states of the unknown covariates $\z$. Why not examine the mean MSE over all possible $\z$ instead of only the worst? Sec.~\ref{deriv:mean_mse} shows that

\bneqn\label{eq:uncond_mean}
\expesubnostr{\z}{\cmsesubnostr{\w}{\betaThat}{\z}} = \oneover{n^2}\f^\top \bSigmaw \f + \overn{\sigsq_z}
\eneqn

if we assume homoskedasticity in the unobserved covariates, i.e. $\var{\z} = \sigsq_z \I_n$ (assumption \setword{A4}{assumption:homoskedasticity}). Note that eliminating the homoskedasticity assumption (while retaining independence of the $z_i$'s) will not substantively change the interpretation of the result which follows.

When taking the $\inf_{\bSigmaw}$ of the above, we minimize the first term that represents imbalance. The second term signifies a fundamental estimation error that cannot be reduced. Thus, the optimal design corresponds to optimal balance, a result noted by \citet[p. 90]{Kallus2018}.

Minimizing the first term under unknown $f$ has been addressed by the same author (ibid, sec. 2.2) who allows nature to choose the response function adversarially in a set of functions $\mathcal{F}$, a normed vector space with norm denoted by $|f|_\mathcal{F}$. Then, he follows another minimax approach, finding the infimum over designs under the supremum of $f \in \mathcal{F}$. Once this supremum is evaluated, the first term of eq.~\ref{eq:uncond_mean} will be an objective function (with the known inputs $\x_1, \ldots, \x_n$) that can be minimized.

Quite shockingly, assumptions about $\mathcal{F}$ imply different well-known objective functions. For example, if $|f|_\mathcal{F}$ was the Lipschitz norm with respect to distance metric $\delta$ between $\x_i$ and $\x_j$, then the objective function to minimize would be the pairwise matching objective with distance $\delta$ (ibid, Theorem 4). If one assumes a linear model, then the objective function would be the Mahalanobis distance between the treatment group average and the control group average (ibid, sec. 2.3.3). Further, he shows that the first term of eq.~\ref{eq:uncond_mean} features an extremely rapid vanishing rate of $O(e^{-n})$ under optimal allocation for parametric response functions (ibid, sec. 3.3). For most objective functions, the problem of finding such an optimal allocation is NP-hard and thus the rate is slower in practice as either approximate polynomial or heuristic methods must be employed.

\subsection{A Tail Criterion}\label{subsec:tail_criterion}

However, considering the average $\z$ may be imprudent when we know the worst case $\z$, is unquestionably ruinous. Moreover, minimizing eq.~\ref{eq:uncond_mean} boils down to minimizing the first term since the second term does not depend on $\bSigmaw$. While the first term can become exponentially small, the second term would remain as $O(1/n)$. This means that the design that minimizes eq.~\ref{eq:uncond_mean} gains very little against other simpler alternatives, e.g. pairwise matching. As we show below, this marginal improvement in the mean criterion comes at the expense of much higher variance yielding a long right tail of the MSE distribution. 

Parenthetically, we note a further weakness of the previous criterion. The unconditional MSE, $\expesubnostr{\z}{\cmsesubnostr{\w}{\betaThat}{\z}}$, is one term in the law of total variance formula. The other term, $\varnostrsub{\z}{\cexpesubnostr{\w}{\betaThat}{\z}}$, is zero since the estimator is unbiased (and thus a constant). This implies that the criterion above of eq.~\ref{eq:uncond_mean} is equivalent to $\varnostrsub{\z,\w}{\betaThat}$. This corresponds to a worldview where replicates consist of both $\w$ and $\z$ being jointly realized together. This is similar to treating $\z$ as \qu{noise} that will be different for each allocation $\w$ for the same subjects. This perspective is substantially at odds with our assumption that $\z$ is affixed to the subjects and is not changed with different allocations.

To remedy both of these problems (but not be as conservative as Efron in assuming the worst case), we propose to optimize the design for the performance of the worst $q$ percent of $\z$'s (i.e. the tail events where $q$ is large),

\bneqn\label{eq:quantile_expression}
\inf_{\bSigmaw}  \braces{\text{Quantile}_{\z}\bracks{\cmsesubnostr{\w}{\betaThat}{\z}, \,q}}.
\eneqn

Note that complete randomization, the result of sec.~\ref{subsec:worse_case_criterion}, can be recovered if we require the worst tail event, i.e. $q = 100\%$.

Once again, we are considering $\cmsesubnostr{\w}{\betaThat}{\z}$ as a random variable in the unobserved covariates. To solve for optimal design, the inverse cumulative distribution function (CDF) must be evaluated at $q$.

However, the expression of eq.~\ref{eq:conditional_mse} is a quadratic form with associated matrix properties that unfortunately are not amenable to current asymptotic distributional results \citep[see e.g.][]{Gotze2002}. If the $z_i$'s were to be normal, the distribution is known, but is not amenable for locating the optimal design (see sec.~\ref{deriv:z_normal_dist}). 

\section{Optimal Designs}\label{sec:optimal_designs}

The worst-case criterion of sec.~\ref{subsec:worse_case_criterion} yields complete randomization as the optimal design. The mean criterion of sec.~\ref{subsec:mean_criterion} yields perfect balance as the optimal design. The criterion that seeks a combination of worst-case and average performance, the tail criterion of sec.~\ref{subsec:tail_criterion}, is less clear. Without the inverse CDF and an explicit means to find the infimum over designs, we cannot provide a procedure to locate the optimal design. 

However, we can prove our eponymous claim, i.e. that the best design is between complete randomization and complete optimization. Denoting $Q$ as the quantile in Expression~\ref{eq:quantile_expression}, note that the quantile expression can be expressed as the mean plus a number of standard errors,

\bneqn\label{eq:hack_criterion}
Q := \expesub{\z}{\cmsesubnostr{\w}{\betaThat}{\z}} + c \times  \sesub{\z}{\cmsesubnostr{\w}{\betaThat}{\z}},
\eneqn

where the quantile of interest $q$ and the number of standard errors $c$ exist in a one-to-one relationship. We are interested in high quantiles, which correspond to large $c$'s. 

The expectation term was found in eq.~\ref{eq:uncond_mean} and the standard error term calculation is found in sec.~\ref{deriv:variance}. Putting both the expectation and standard error together we have

\bneqn\label{eq:hack_criterion_official}
Q = \oneover{n^2} \Big(
\underbrace{\f^\top \bSigmaw \f}_{B_1} 
\ingray{ + n\sigsq_z} + c 
\big(
\ingray{n \kappa_z} + 
2 \sigsqzsq \underbrace{\frobsq{\bSigmaw}}_{R} \,+\, 
4 \sigsq_z \underbrace{\f^\top \bSigmaw^2 \f}_{B_2}
\big)^{\half}
\Big)
\eneqn

where $\kappa_z := \expe{z^4} - 3(\sigsq_z)^2$, which is zero if $\z$ is normally distributed and $\frobsq{\bSigmaw}$ is the squared Frobenius norm of $\bSigmaw$. To compute $Q$, we both need to assume $\z$ has a finite fourth moment (assumption \setword{A5}{assumption:response_fourth_moment}) and that its third and fourth moments are independent of $\X$ (assumption \setword{A6}{assumption:third_fourth_moments_independent}). The two gray terms are constants unaffected by our choice of design. There is another term in the standard error that is eliminated via assuming forced balance designs between the number of treatments and controls (assumption \setword{A7}{assumption:forced_balance}).

There is a further complication. Given a desired quantile $q$, the constant $c$ will be a function of $\bSigmaw$ (although we will see in sec.~\ref{sec:simulations} that it changes only slightly). There are a couple of default values of $c$ we can employ. First, we can use the Chebyshev constant, $1 / \sqrt{1 - q}$, thus our criterion becomes a bound on the quantile and not the quantile itself. Second, we can unconditionally employ $c = 2$ as in the Gaussian setting, a value that is shown to be approximately true for a large set of continuous distributions \citep{Sharpe1970}. Any choice of $c$ here will not affect the asymptotic argument we now turn to.

To prove our claim, we asymptotically examine the three terms affected by the design, denoted $B_1,\, B_2$ and $R$ as the first two measure imbalance in the portion of the response that is observed and the last loosely measures the degree of randomness, serving as a \qu{regularization} term in a sense as it \qu{shrinks} the design back towards randomization. We do this analysis for complete randomization and for perfect optimization.

For complete randomization with forced balance, $\bSigmaw = \frac{n}{n-1}\I_n-\frac{1}{n-1}\onevec_n \onevec_n^\top$ and thus the $B_1$ and $B_2$ terms are equal to $\normsq{\f} (1+O(1/n))= O(n)$ because $\normsq{\f} = O(n)$ and the $R$ term is $n+\frac{n}{n-1}$ corresponding to

\beqn
Q_{CRFB} = \oneover{n^2} 
\Big(
\underbrace{O(n)}_{B_1}
\ingray{ + O(n)} + 
(\ingray{O(n)} + 
\underbrace{O(n)}_{R} + 
\underbrace{O(n)}_{B_2}
)^{\half}\Big).
\eeqn

For near perfect balance (PB) via optimization, the resultant $\bSigmaw = \w_* \w_*^\top$ (where $\w_*$ is the optimal allocation) and has -1's and +1's in the off-diagonal. The $R$ term is exactly $n^2$ since $\bSigmaw$ has only one non-zero eigenvalue: $n$. The asymptotic analysis of the $B_1$ and $B_2$ terms uses the result from \citet[sec. 3.3]{Kallus2018} twice. First, the $B_1$ term is $O(e^{-n})$. Second, the $B_2$ term is bounded by $n \f^\top \bSigmaw \f$ so it behaves at most like $O(ne^{-n})$. Thus, 

\beqn
Q_{PB} = \oneover{n^2} 
\Big(
\underbrace{O(e^{-n})}_{B_1} 
\ingray{ +O(n)} + 
(\ingray{O(n)} + 
\underbrace{O(n^2)}_{R} + 
\underbrace{O(ne^{-n})}_{B_2}
)^{\half}\Big).
\eeqn

When considering both analyses, the two balance terms vary between order of $n$ down to $e^{-n}$, but the randomness term varies between order of $n^{1/2}$ to $n$. Therefore, while the perfect optimization obtains very small balance terms $B_1$ and $B_2$, the gain is undone by the large randomness term $R$. We conclude that the optimal design must feature balance terms with order slower than $O(e^{-n})$ to guarantee the randomness term does not grow as fast as $O(n)$. This design must be between complete randomization and perfect balance. 

Note that the constant that figures most prominently into our tail criterion of eq.~\ref{eq:hack_criterion_official} is $\sigsq_z$. Smaller values indicate that the observed covariates explain more of the spread in the response (i.e. $R^2$ is larger).

\section{Simulations}\label{sec:simulations}

In the model of eq.~\ref{eq:main_model}, we assume $n=20$, one covariate and let $\f = \x$ (the simplest case) and $\beta_T = 1$. We generate one fixed $\x$ whose values are \emph{iid} realizations from $\stdnormnot$. This simulation exercise is conditional on those values. We then generate all forced balance vectors in $\allocspace$ (i.e. $\binom{n}{n/2} \approx 3.7$ million $\w$ vectors). 

Since the model is linear, the appropriate imbalance function to be minimized is the Mahalanobis distance between the average observed covariates in the treatment group and the average of the observed covariates in the control group. With one covariate, the imbalance objective reduces to $\abss{\bar{x}_T - \bar{x}_C}$. The optimal vector $\w_*$ is then located and saved.

We consider three designs: (CRFB) complete randomization with forced balance (PB) perfect balance using $\w_*$ and (PM) the pairwise matching algorithm of \citet{Greevy2004}. The third is to demonstrate a design between complete randomization and perfect balance. The choice of matching does not suggest that we view it as the true optimal design; it is merely an illustration. To do the pairwise matching, the $x_i$'s are sorted and each set of two becomes a pair. 

In each iteration we first draw one $\z$ whose entries are \emph{iid} $\normnot{0}{\sigsq_z}$ where the choice of variance was made so that $R_x^2$ of the model was approximately 35\% for the observed covariates ($\sigsq_z = 1.5^2$). We then repeat the following 2,000 times for each design. 

(CRFB) We choose at random 300 $\w$'s from $\allocspace$ to simulate the CRFB design. For each, we compute $\y$ and then $\betaThat$. Over these 300, we estimate $\cmsesubnostr{\w}{\betaThat}{\z}$ by taking the sample average $(\betaThat - \beta_T)^2$ over these 300. (PB) We compute $\y$ for the allocations $\w_*$ and $-\w_*$ and then $\betaThat$ for both. The average of both $(\betaThat - \beta_T)^2$ values is the estimate of $\cmsesubnostr{\w}{\betaThat}{\z}$. (PM) We choose at random 300 $\w$'s by randomly permuting $+1, -1$ within the 10 pairs of subjects. The procedure for CRFB is then repeated.

\begin{figure}[htp]
\centering
\includegraphics[width=6.5in]{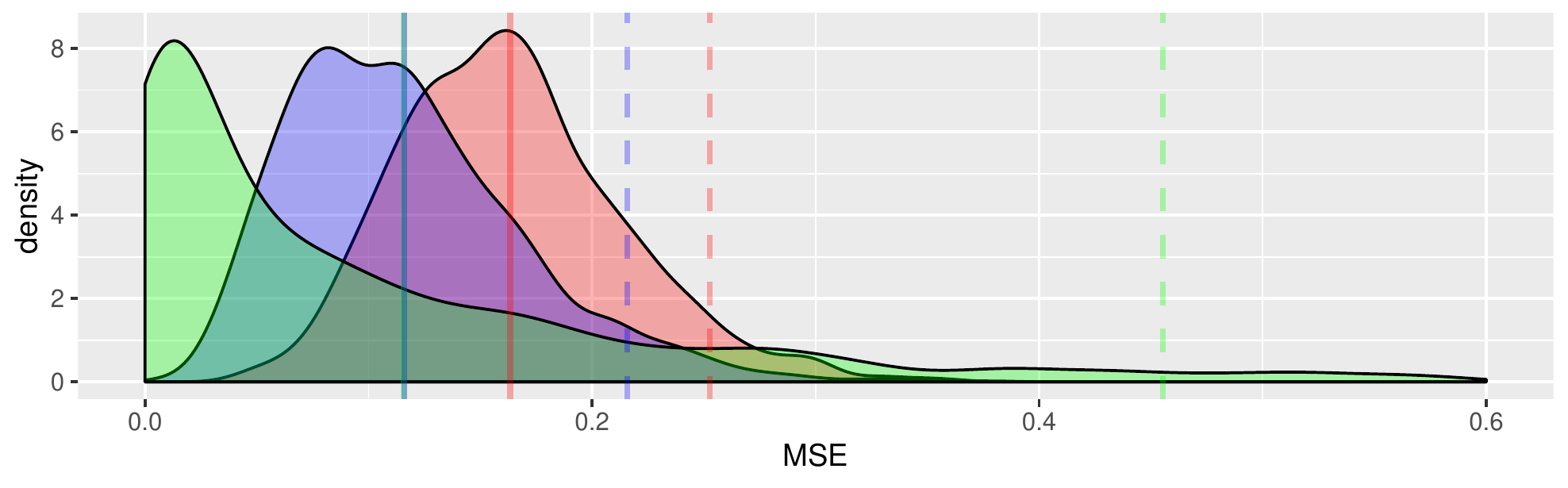}
\caption{Density estimates for the MSE distribution for CRFB (red), perfect balance (green) and pairwise matching (blue) over the 2,000 simulation replicates explained in the text. Solid vertical lines are the estimated mean MSE and dashed vertical lines are the estimated 95\% percentile of the MSE. Note that the mean MSE for perfect balance and pairwise matching are nearly identical with perfect balance ahead by a small margin because the gains in over-optimizing the $B_1$ term are slim.}
\label{fig:sim_results_Rsq_36_percent}
\end{figure}

Fig.~\ref{fig:sim_results_Rsq_36_percent} illustrates for all three designs the distribution of $\cmsesubnostr{\w}{\betaThat}{\z}$ over the 2,000 draws from the $\z$ distribution. 

There are many observations from this illustration. First, the solid vertical lines indicate the estimated $\expesubnostr{\z}{\cmsesubnostr{\w}{\betaThat}{\z}}$, the mean criterion of sec.~\ref{subsec:mean_criterion}. Here, the perfect balance design is optimal (matching is a close runner-up) as explained in the beginning of sec.~\ref{sec:optimal_designs}. Second, the worst-case criterion of sec.~\ref{subsec:worse_case_criterion} is assessed by the maximum values which are CRFB: 0.36, PB: 1.65 (note the figure only shows up to 0.6!), PM: 0.38. This value is smallest for the CRFB design as explained in the beginning of sec.~\ref{sec:optimal_designs}. Last, the dashed vertical lines illustrate the estimated $\text{Quantile}_{\z} [ \cmsesubnostr{\w}{\betaThat}{\z}, \,95\% ]$, our recommended criterion of sec.~\ref{subsec:tail_criterion}. Here, perfect balance performs terribly as CRFB is clearly more optimal. Pairwise matching is an example design that is between these two extremes and beats both of them. However the true optimal, an elusive harmony between randomization and optimization, will perform better than even matching as the asymptotic analyses of sec.~\ref{sec:optimal_designs} demonstrates. This optimal design is \qu{closer} to CRFB than to PB.

Additionally, there is the concern of the constant $c$ in eq.~\ref{eq:hack_criterion_official} varying with fixed $q =95\%$. In the above simulation, the values of $c$ for CRFB, PM and PB are 1.75, 1.84 and 1.85. If $c$ were set to be 2, the story would not differ.

\begin{figure}[htp]
\centering
\includegraphics[width=6.5in]{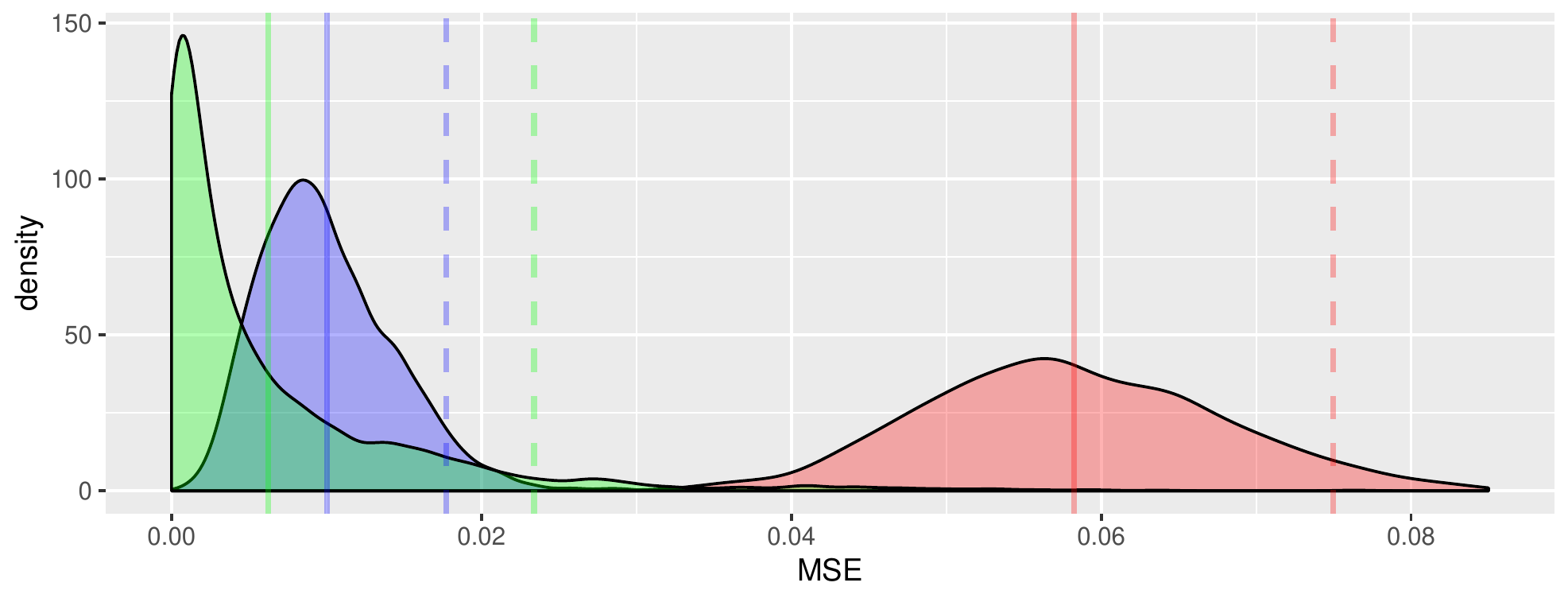}
\caption{Same as fig.~\ref{fig:sim_results_Rsq_36_percent} except the observed covariates and PATE account for the 90\% of the variance in the response.}
\label{fig:sim_results_Rsq_90}
\end{figure}

We also demonstrate two more extreme examples. The first is where there is no effect of observed covariates (accomplished by allowing $\f = \zerovec_n$). The second is where $R_z^2 = 5\%$, i.e. the unobserved covariates contribute very little to the response (accomplished by setting $\sigsq_z = 0.01^2$).  and fig.~\ref{fig:sim_results_Rsq_90} illustrate the results.

\begin{figure}[htp]
\centering
\includegraphics[width=6.5in]{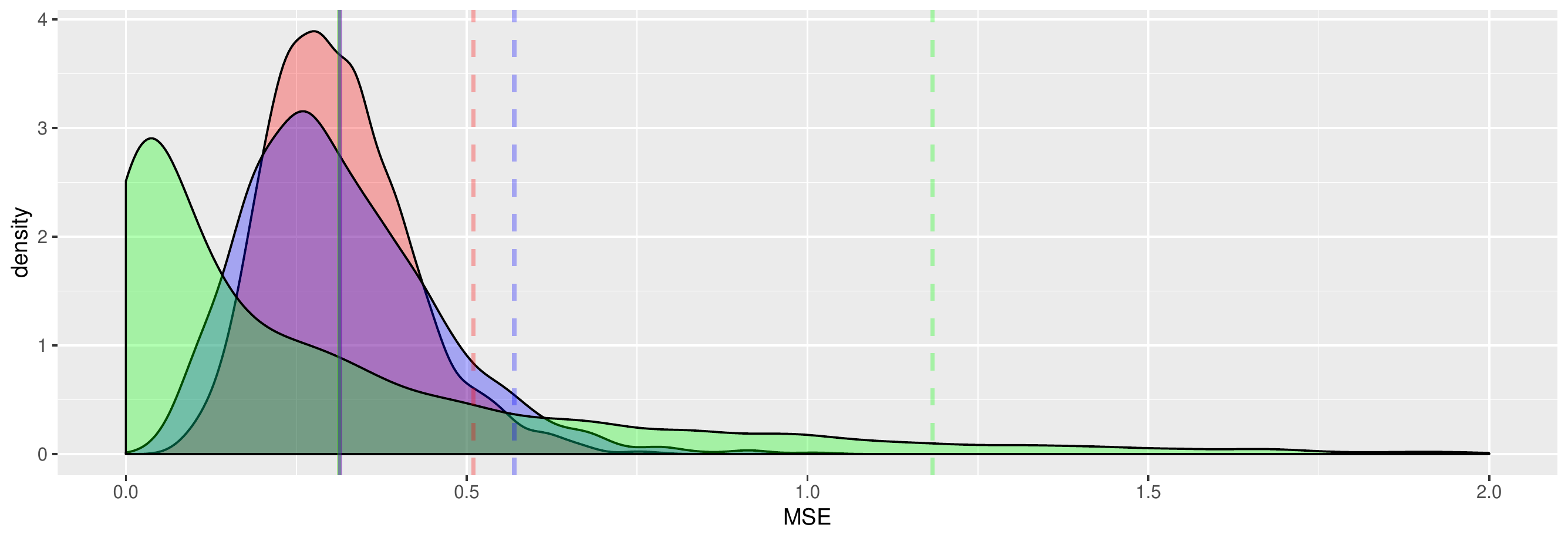}
\caption{Same as fig.~\ref{fig:sim_results_Rsq_36_percent} except the observed covariates have no effect on the response.}
\label{fig:sim_results_Rsq_0}
\end{figure}

Fig.~\ref{fig:sim_results_Rsq_0} illustrates the case of no effect of the observed covariates. Here, the mean MSE is the same for PB, PM and CRFB procedures. This is expected as the balance term ($B_1$) in eq.~\ref{eq:uncond_mean} is 0 (since $\f = \zerovec_n$) hence the mean MSE is $\overn{\sigsq_z}$ regardless of the procedure. However, the 95\% quantile of MSE is minimized with CRFB. This is because any restriction on the allocation will increase the value of the $R$ term in eq.~\ref{eq:hack_criterion_official} with no corresponding decrease in the $B_1$ and $B_2$ terms and hence CRFB is optimal here.

Fig.~\ref{fig:sim_results_Rsq_90} illustrates  the case where the observed covariates are the significant driver of the response. We note that optimal design still lies between CRFB and PB when considering case of the 95\%ile tail criterion (as in fig.~\ref{fig:sim_results_Rsq_36_percent}). In contrast (a) the mean criterion more clearly illustrates thet dominance of using PB designs over randomized designs and (b) the optimal design, still a harmony of PB and CRFB, is now \qu{closer} to PB than to CRFB (since the small $\sigsq_z$ makes the $B_1$ term dominate over the $R$ term in eq.~\ref{eq:hack_criterion_official}).

Taking the last scenario to its limit where the unobserved covariates are omitted, we will find that optimization is the best design; and randomization is inferior by a large margin. This can be seen in eq.~\ref{eq:hack_criterion_official}. Since $\sigsq_z \approx 0$, there is no effect of any term except the $B_1$ balance term.

In our final simulation, we duplicate the first scenario with $n=200$ to illustrate our titular claim maintains in sample sizes common in real-world studies. There is one complication. When $n$ was 20, the optimal $\w_*$ was found by brute force; at $n=200$, this is no longer possible. To approximate $\w_*$, we use the numerical method of \citet{Krieger2016}. Their algorithm switches pairs of subjects greedily to minimize observed imbalance until a local optimum is found. Using the \texttt{R} package \texttt{GreedyExperimentalDesign}, we repeat their algorithm 20,000 times and find the $\w$ with the minimum imbalance. Here, $(\bar{x}_T - \bar{x}_C)^2$ is on the order of $10^{-22}$ and matching is on the order of $10^{-5}$. 

\begin{figure}[htp]
\centering
\includegraphics[width=6.5in]{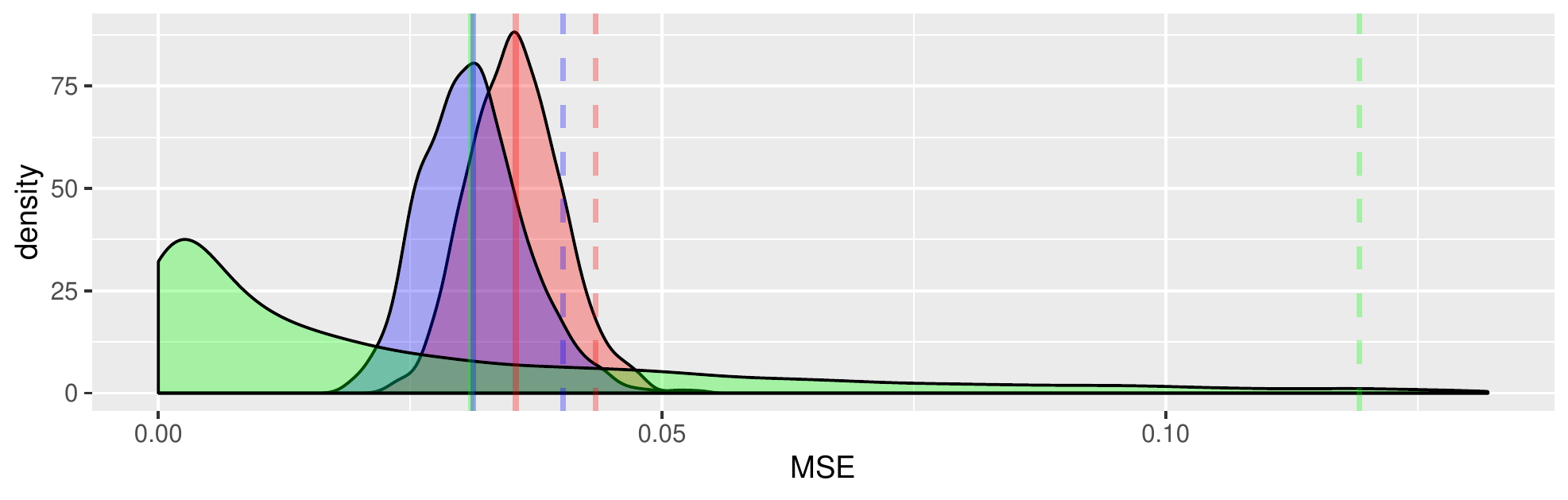}
\caption{Same as fig.~\ref{fig:sim_results_Rsq_36_percent} except the the sample size is now $n=200$.}
\label{fig:high_n_results}
\end{figure}

The results are displayed in fig.~\ref{fig:high_n_results} and we observe the same results as previously. The results for the optimal design were robust to different approximations to $\w_*$ ranging from $10^{-14}$ up to $10^{-22}$. For the true optimal (which is impossible to find), the results will be very similar. Also, the values of $c$ for CRFB, PM and PB were 1.72, 1.76 and 2.04. Again, if $c$ were set to be 2, the story would not differ. 

\section{Discussion}

We assume a two-arm randomized fixed trial with a response model that can be decomposed into a sum of an observed measurement component, a treatment effect and an unobserved measurement component. Using the differences-in-mean estimator and worrying about a tail of extreme events in its mean-squared error, we have shown that CRFB is too conservative and optimizing the observed measurements' balance is too aggressive. The optimal design lies somewhere between these two extremes.

To create an algorithm to find this optimal design will require an explicit minimum of eq.~\ref{eq:hack_criterion_official}. Since the response function $f$ is unknown, a supremum over its worst case is likely prudent. The best design will also depend on constants that must be estimated a priori. The most important being the proportion of variance in the response explained by the observed covariates. The less the covariates matter in this respect, the closer the optimal design will be to CRFB; the more they matter, the closer the optimal design will be to optimization \footnote{Even if the optimal variance-covariance matrix can be computed, it is known that this matrix does not fully specify $\prob{\w}$ or $\allocspace_0$ \citep{Teugels1990}. Thus, we would need to employ a numeric algorithm to draw $\w$'s whose entries have the correct covariances.}. 

Our criterion suggested in this paper is a quantile of the distribution of the mean squared error of the mean-differences estimator. A common alternate estimator in practice is the ordinary least squares regression estimator for $\beta_T$. The mathematics herein can be redone for this estimator. Unshown simulations reveal the same story as in Figures~\ref{fig:sim_results_Rsq_36_percent}--\ref{fig:high_n_results}. 

Additionally, reductions in MSE of the estimator are usually only important insofar as they increase power. More work needs to be done with power simulations. This brings up hypothesis testing, a topic avoided in this work. Hypothesis testing is complicated in restricted randomization as the finite distribution of the estimator are conditional upon the randomization procedure.

Given the simulation study herein and an asymptotic analysis, complete randomization with forced balance is still likely the best policy pending further necessary research.

\if0\blind{
\subsection*{Acknowledgement}

We thank Michael Sklar and Bracha Blau  for helpful discussions.
}\fi

\bigskip
\begin{center}
{\large\bf SUPPLEMENTARY MATERIAL}
\end{center}

\input{appendices}

\bibliographystyle{Chicago}
\bibliography{refs}

\end{document}

%% file: preamble.tex
\usepackage{graphicx}
\usepackage{subcaption}
\usepackage{amsmath}
\usepackage{dsfont}
\usepackage{amssymb}
\usepackage{abstract}
\usepackage[auth-sc,affil-sl]{authblk}
\usepackage{hyperref}
\usepackage{parskip}
\usepackage{enumerate}
\usepackage{fancyhdr}
\usepackage[xcdraw,table]{xcolor}
\usepackage{algorithm}
\usepackage{algorithmicx}
\usepackage{algpseudocode}
\usepackage[table]{xcolor}
\usepackage{multirow}
\usepackage{amsthm}

\usepackage{fancyvrb}

\newcounter{probnum}
\allowdisplaybreaks

\definecolor{tabblue}{rgb}{.870588,.905882,.94902}
\definecolor{gray}{rgb}{0.7,0.7,0.7}
\definecolor{black}{rgb}{0,0,0}
\definecolor{white}{rgb}{1,1,1}
\definecolor{blue}{rgb}{0.0,0.0,1}
\newcommand{\ingray}[1]{\color{gray}#1\color{black}}

\definecolor{green}{rgb}{0,0.5,0}

\definecolor{yellow}{rgb}{1,0.549,0}

\definecolor{red}{rgb}{0.6,0.0,0.0}

\definecolor{darkred}{rgb}{0.9,0.4,0}

\definecolor{purple}{rgb}{0.58,0,0.827}

\definecolor{backgcode}{rgb}{0.97,0.97,0.8}
\definecolor{Brown}{cmyk}{0,0.81,1,0.60}
\definecolor{OliveGreen}{cmyk}{0.64,0,0.95,0.40}
\definecolor{CadetBlue}{cmyk}{0.62,0.57,0.23,0}

\newcommand{\qu}[1]{``{#1}''}
\newcommand{\bv}[1]{\boldsymbol{#1}}

\newcommand{\sigsq}{\sigma^2}

\newcommand{\sigsqsq}{\parens{\sigma^2}^2}
\newcommand{\sigsqzsq}{\parens{\sigma_z^2}^2}

\newcommand{\bSigmaw}{\bv{\Sigma}_{\bf w}}

\newcommand{\betaThat}{\hat{\beta}_T}

\newcommand{\allocspace}{\mathcal{W}}

\newcommand{\half}{\frac{1}{2}}

\newcommand{\X}{\bv{X}}

\newcommand{\I}{\bv{I}}

\newcommand{\x}{\bv{x}}
\newcommand{\f}{\bv{f}}
\newcommand{\w}{\bv{w}}
\newcommand{\onevec}{\bv{1}}
\newcommand{\zerovec}{\bv{0}}

\newcommand{\y}{\bv{y}}

\renewcommand{\v}{\bv{v}}

\newcommand{\z}{\bv{z}}

\newcommand{\reals}{\mathbb{R}}

\newcommand{\beqn}{\vspace{-0.25cm}\begin{eqnarray*}}
\newcommand{\eeqn}{\end{eqnarray*}}
\newcommand{\bneqn}{\vspace{-0.25cm}\begin{eqnarray}}
\newcommand{\eneqn}{\end{eqnarray}}
\newcommand{\benum}{\begin{enumerate}}
\newcommand{\eenum}{\end{enumerate}}

\newcommand{\parens}[1]{\left(#1\right)}
\newcommand{\squared}[1]{\parens{#1}^2}

\newcommand{\prob}[1]{\mathbb{P}\parens{#1}}

\newcommand{\sumonen}[2]{\sum_{#1=1}^n #2}

\newcommand{\bracks}[1]{\left[#1\right]}
\newcommand{\braces}[1]{\left\{#1\right\}}

\newcommand{\abss}[1]{\left|#1\right|}
\newcommand{\norm}[1]{\left|\left|#1\right|\right|}
\newcommand{\normsq}[1]{\norm{#1}^2}
\newcommand{\frobsq}[1]{\normsq{#1}_F}

\newcommand{\tr}[1]{\text{tr}\bracks{#1}}

\newcommand{\expe}[1]{\mathbb{E}\bracks{#1}}
\newcommand{\cexpenostr}[2]{\mathbb{E}[#1\,|\,#2]}
\newcommand{\cexpe}[2]{\expe{#1\,|\,#2}}

\newcommand{\cexpesubnostr}[3]{\mathbb{E}_{\,#1}[#2\,|\,#3]}

\newcommand{\cvarnostr}[2]{\varnostr{#1\,|\,#2}}

\newcommand{\expesub}[2]{\mathbb{E}_{\,#1}\bracks{#2}}
\newcommand{\expesubnostr}[2]{\mathbb{E}_{\,#1}[#2]}

\newcommand{\var}[1]{\mathbb{V}\text{ar}\bracks{#1}}
\newcommand{\varsub}[2]{\mathbb{V}\text{ar}_{#1}\bracks{#2}}
\newcommand{\varnostr}[1]{\mathbb{V}\text{ar}[#1]}
\newcommand{\varnostrsub}[2]{\mathbb{V}\text{ar}_{#1}[#2]}

\newcommand{\cmsesubnostr}[3]{\mathbb{M}\text{SE}_{#1}[#2\,|\,#3]}

\newcommand{\sesub}[2]{\mathbb{S}\text{E}_{#1}\bracks{#2}}

\newcommand{\oneover}[1]{\frac{1}{#1}}
\newcommand{\overtwo}[1]{\frac{#1}{2}}
\newcommand{\overn}[1]{\frac{#1}{n}}

\newcommand{\multnormnot}[3]{\mathcal{N}_{#1}\parens{#2,\,#3}}

\newcommand{\normnot}[2]{\mathcal{N}\parens{#1,\,#2}}

\newcommand{\stdnormnot}{\normnot{0}{1}}

\newcommand{\mult}[1]{\text{Mult}\bracks{#1}}
\newcommand{\ncchisq}[2]{\chi^{`2}\parens{#1,\,#2}}

\newcommand{\betaT}{\beta_T}


%% file: appendices.tex
\subsection*{Replication}

All figures and tables can be reproduced by running the \texttt{R} code found at
\if0\blind{
\url{https://github.com/kapelner/harmonizing_designs/blob/master/paper_duplication.R}.
}\fi
\if1\blind{
\url{https://github.com/[blinded]/blob/master/paper_duplication.R}.
}\fi

\section{APPENDIX: Proofs}

\subsection{Proof of Unbiasedness}\label{pf:unbiasedness}

We now show that $\expesubnostr{\w,\z}{\betaThat}=\beta_T$. The law of iterated expectation implies that $\expesubnostr{\w,\z}{\betaThat} = \expesubnostr{\z}{\cexpesubnostr{\w}{\betaThat}{\z}}$. We solve the inner expecation below using the model given by Equation~\ref{eq:main_model}:

\bneqn
\cexpesubnostr{\w}{\betaThat}{\z} &=& \oneover{n} \cexpesubnostr{\w}{\w^\top \parens{\betaT \w + \f + \z}}{\z} \\
&=& \oneover{n} \parens{
\cexpesubnostr{\w}{\betaT \w^\top \w}{\z} +
\cexpesubnostr{\w}{\w^\top \f}{\z} +
\cexpesubnostr{\w}{\w^\top \z}{\z}
} \\
&=& \oneover{n} \parens{
\betaT \cexpesubnostr{\w}{\w^\top \w}{\z} +
\cexpesubnostr{\w}{\w^\top \f}{\z} +
\expesubnostr{\w}{\w}^\top \z
}
\eneqn

Since $w_i \in \{-1,+1\}$ then, ${\w^\top\w}=\sum_{i=1}^n w_i^2 =n$. The mirror property (A2) implies that $\expe{\w^\top\f}=0$, since

\beqn
\expesub{\w}{\w^\top \f}=\sum_{\w_0 \in \allocspace_0}\w_0^T\f \,\prob{\w=\w_0}
\eeqn

for all $\allocspace_0$ that we consider. Then by \ref{assumption:mirror}, each summand corresponding to certain $\w_0$ cancels out with the summand with $-\w_0$. Therefore, $\expesub{\w}{\w^\top \f}=0$.  

Also by the mirror property, $\expesubnostr{\w}{\w} = \zerovec_n$. This leaves us with:

\beqn
\cexpesubnostr{\w}{\betaThat}{\z} &=& \betaT
\eeqn

The unconditional expectation is equivalent.

\subsection{Derivation of the Conditional MSE}\label{deriv:cond_var}

The unbiasedness of $\betaThat~|~\z$ (proven in sec.~\ref{pf:unbiasedness}) implies that $\cvarnostr{\betaThat}{\z} = \cexpenostr{\betaThat^2}{\z} - \beta_T^2$ where the expectation is taken over $\w \in \mathcal{W}_0$. Recall that $\betaThat := \w^\top \y /n$ so that

\beqn
\cvarnostr{\betaThat}{\z} &=& \cexpenostr{(\w^\top \y /n)^2}{\z} \\
&=& \oneover{n^2} \expe{
\squared{
\w^\top (\beta_T \w + \f + \z)
}
\,\Big|\, \z} - \beta_T^2\\
&=& \oneover{n^2} \expe{
\squared{
\beta_T \w^\top \w + \w^\top(\f + \z)
}
\,\Big|\, \z} - \beta_T^2
\eeqn

Note that  $\w^\top \w=\sum_{i=1}^n w_i^2 = n$, since $w_i \in \{-1,+1\}$. After canceling out the constant $\beta_T^2$, we are left with:

\beqn
&=& \oneover{n^2} \expe{
2n\beta_T \w^\top (\f + \z) +
(\w^\top(\f + \z))^2
\,\Big|\, \z}
\eeqn

By the same arguments of sec.~\ref{pf:unbiasedness}, $\expesub{\w}{\w^\top (\f + \z)} = 0$ leaving us with

\beqn
&=& \oneover{n^2} \expe{
(\w^\top(\f + \z))^2
\,\Big|\, \z} =
\oneover{n^2} \expe{
(\f + \z)^\top \w \w^\top (\f + \z)
\,\Big|\, \z} = 
\oneover{n^2} 
(\f + \z)^\top \bSigmaw (\f + \z).
\eeqn

\subsection{Derivation of the Mean MSE}\label{deriv:mean_mse}

We wish to find $\expesubnostr{\z}{\cmsesubnostr{\w}{\betaThat}{\z}}$. Using the result from sec.~\ref{deriv:cond_var},

\beqn
\expesubnostr{\z}{\cmsesubnostr{\w}{\betaThat}{\z}} &=& \expesub{\z}{\oneover{n^2}(\f + \z)^\top \bSigmaw (\f + \z)} \\
&=& \oneover{n^2}  \expesub{\z}{\f^\top \bSigmaw \f + 2 \f^\top \bSigmaw \z + \z^\top \bSigmaw \z} \\
&=& \oneover{n^2}\f^\top \bSigmaw \f + \frac{2}{n^2} \f^\top \bSigmaw \expesub{\z}{\z} + \oneover{n^2}\expesub{\z}{\z^\top \bSigmaw \z} \\
\eeqn 

Since $\expesubnostr{\z}{\z} = \zerovec_n$ by construction, the second term is zero. 

The third term is the expectation of a quadratic form that is the trace of the associated matrix times the variance-covariance matrix of the vector plus the quadratic form of the expectation vector and the associated matrix. Since $\expe{\z} = \zerovec_n$ (by \ref{assumption:additivity}), we only need to consider the first term. Assuming homoskedasticity (\ref{assumption:homoskedasticity}), $\var{\z} = \sigsq_z \I_n$ and $\tr{\bSigmaw} = n$, the expression evaluates to $n\sigsq_z$ since the diagonal entries are identifically one. The heteroskedastic case will not be substantively different. Thus,

\beqn
\expesubnostr{\z}{\cmsesubnostr{\w}{\betaThat}{\z}} &=& \oneover{n^2}\f^\top \bSigmaw \f + \overn{\sigsq_z}.
\eeqn 

\subsection{Distribution of the MSE Under Normality}\label{deriv:z_normal_dist}

If we assume $\z \sim \multnormnot{n}{\zerovec_n}{\sigsq_z \I_n}$, then $\f + \z \sim \multnormnot{n}{\f}{\sigsq_z \I_n}$. We now examine the distribution of the quadratic form, $(\f + \z)^\top \bSigmaw (\f  + \z)$ where $\bSigmaw$ has properties outlined in the text. \citet{Baldessari1967} proves that this quadratic form is distributed as

\beqn
(\f + \z)^\top \bSigmaw (\f  + \z) \sim  \sum_{i=1}^s \sigsq_z \lambda_j \ncchisq{\mult{\lambda_i}}{\overtwo{\sigsq_z} \f^\top \v_i \v_i^\top \f}
\eeqn

where $\lambda_1 \geq \ldots \geq \lambda_n \geq 0$ and the $\v_i$'s are the unique eigenvalues and eigenvectors of $\bSigmaw$, $\mult{\lambda_i}$ is the multiplicity of the eigenvalue $\lambda_i$ and $\ncchisq{\nu}{\lambda}$ denotes a non-central $\chi^2$ random variable with  degrees of freedom $\nu$ and non centrality parameter $\lambda$. Thus, it is a sum of scaled non-central $\chi^2$ random variables.

Since the distribution is parameterized by the eigendecomposition of $\bSigmaw$, it would be very difficult to optimize the inverse CDF over the space of legal matrices.

\subsection{Derivation of the Variance of the MSE}\label{deriv:variance}

We wish to derive an expression for the variance of the expected squared loss function where the expectation is taken over all randomizations and the variance is taken over all unobserved covariates realizations, i.e. $\varsub{\z}{\cmsesubnostr{\w}{\betaThat}{\z}}$.

From sec.~\ref{deriv:cond_var}, we learned that $\cmsesubnostr{\w}{\betaThat}{\z} =\oneover{n^2} (\f + \z)^\top \bSigmaw (\f + \z)$. This is a variance of a quadratic form, which can be calculated via \citet[eq. 319]{matrixcookbook} when assuming that the conditional third and fourth moment of $\z$ are finite (\ref{assumption:response_fourth_moment}) and do not depend on $\x$ (\ref{assumption:third_fourth_moments_independent}). Thus we have,

\bneqn\label{eq:mse_as_function_of_z}
\varsub{\z}{\cmsesubnostr{\w}{\betaThat}{\z}} = \oneover{n^4} 
\parens{
	n \kappa_z + 2 \sigsqzsq \frobsq{\bSigmaw} + 4 \sigsq_z \f^\top \bSigmaw^2 \f+\gamma_z \onevec_n^\top \bSigmaw \f
},
\eneqn

where  $\kappa_z := \expe{z^4} - 3\sigsqsq$ and $\gamma_z := \expe{z^3}$. We prove in sec.~\ref{app:skewness_term_zero} that the last expression in eq.~\ref{eq:mse_as_function_of_z} above is zero and therefore eq.~\ref{eq:hack_criterion_official} follows.

\subsection{A Proof that the Last MSE Term is Zero}\label{app:skewness_term_zero}

We wish to demonstrate that $\onevec_n^\top \bSigmaw \f = \f^\top \bSigmaw \onevec_n = 0$. First, we reiterate that $\onevec_n^\top \w = 0$  by the assumption of forced balance (\ref{assumption:forced_balance}). This also means that $\expesubnostr{\w}{\onevec_n^\top \w} = \varnostrsub{\w}{\onevec_n^\top \w} = 0$ since every $\w$ is balanced. Then, $\varnostrsub{\w}{\onevec_n^\top \w} = \expesubnostr{\w}{\squared{\onevec_n^\top \w}} = \onevec_n^\top \bSigmaw \onevec_n = 0$.

Note that $\bSigmaw = \sumonen{i}{\lambda_i \v_i \v_i^\top}$ where the $\lambda_1 \geq \ldots \geq \lambda_n \geq 0$ and $\v_i$'s are its eigenvalues and eigenvectors respectively. Since $\bSigmaw$ is a variance-covariance matrix, it is symmetric implying that its eigenvalues are all non-negative. We can then write $\onevec_n^\top \bSigmaw \onevec_n = \sumonen{i}{\lambda_i \squared{\onevec_n^\top \v_i}} = 0$. This means that $\lambda_i \squared{\v_i^\top \onevec_n} = 0$ for all $i$. In order for this to be true, for every $i$ either $\lambda_i = 0$ or $\v_i^\top \onevec_n = 0$.

We now examine just the term $\bSigmaw \onevec_n$ which can be written as $\sumonen{i}{\lambda_i \v_i \v_i^\top \onevec_n}$. For all $i$ either $\lambda_i$ or $\v_i^\top \onevec_n$ is zero rendering the \qu{middle} $\v_i$ irrelevant. Thus $\bSigmaw \onevec_n = \zerovec_n$ and $\f^\top \bSigmaw \onevec_n = \f^\top \zerovec_n = 0$.